\documentclass[a4paper,twoside]{article}

\usepackage{epsfig}
\usepackage{hyperref}
\usepackage{subcaption}
\usepackage{calc}
\usepackage{amssymb}
\usepackage{amstext}
\usepackage{amsmath}
\usepackage{amsthm}
\usepackage{multicol}
\usepackage{pslatex}
\usepackage{apalike}
\usepackage{algorithm2e}
\usepackage{array}
\usepackage{listings}
\usepackage{tcolorbox}
\usepackage[bottom]{footmisc}
\usepackage{graphicx} 
\usepackage{adjustbox} 
\usepackage{SCITEPRESS}
\usepackage{tikz}
\usetikzlibrary{trees}
\usepackage{xcolor}
\definecolor{codeBackground}{rgb}{0.95,0.95,0.92} 

\lstdefinelanguage{json}{
    basicstyle=\ttfamily\small,
    numbers=left,
    numberstyle=\tiny,
    stepnumber=1,
    numbersep=8pt,
    showstringspaces=false,
    breaklines=true,
    frame=lines,
    backgroundcolor=\color{codeBackground}, 
    morestring=[b]",
    stringstyle=\color{blue},
    commentstyle=\color{gray},
    keywordstyle=\color{purple},
    morekeywords={true,false,null} 
}

\begin{document}

\title{LLM-Generated Microservice Implementations from RESTful API Definitions}

\author{
    \authorname{
        Saurabh Chauhan\sup{1}, Zeeshan Rasheed\sup{1}, Abdul Malik Sami\sup{1}, Zheying Zhang\sup{1},Jussi Rasku\sup{1}, Kai-Kristian Kemell\sup{1}, and Pekka Abrahamsson\sup{1}
    }
    \affiliation{\sup{1}Faculty of Information Technology and Communication Science, Tampere University, Finland}
    \email{\{saurabh.chauhan, zeeshan.rasheed, malik.sami, zheying.zhang, jussi.rasku, kai-kristian.kemell, pekka.abrahamsson\}@tuni.fi}
}


\abstract{The growing need for scalable, maintainable, and fast-deploying systems has made microservice architecture widely popular in software development. This paper presents a system that uses Large Language Models (LLMs) to automate the API-first development of RESTful microservices. This system assists in creating OpenAPI specification, generating server code from it, and refining the code through a feedback loop that analyzes execution logs and error messages.
By focusing on the API-first methodology, this system ensures that microservices are designed with well-defined interfaces, promoting consistency and reliability across the development life-cycle. The integration of log analysis enables the LLM to detect and address issues efficiently, reducing the number of iterations required to produce functional and robust services. This process automates the generation of microservices and also simplifies the debugging and refinement phases, allowing developers to focus on higher-level design and integration tasks. This system has the potential to benefit software developers, architects, and organizations to speed up software development cycles and reducing manual effort. To assess the potential of the system, we conducted surveys with six industry practitioners. After surveying practitioners, the system demonstrated notable advantages in enhancing development speed, automating repetitive tasks, and simplifying the prototyping process. While experienced developers appreciated its efficiency for specific tasks, some expressed concerns about its limitations in handling advanced customizations and larger-scale projects. The code is publicly available at \url{https://github.com/sirbh/code-gen} }

\onecolumn \maketitle \normalsize \setcounter{footnote}{0} \vfill

\section{\uppercase{Introduction}}
\label{sec1}

There is growing interest in microservice architecture among organizations \cite{10363466}. This approach is getting popular because it breaks down application components into smaller independent services, which makes it easier to scale, develop, and deploy \cite{8305969}. Each service in a microservices architecture is built independently to fulfill a specific function, enabling focused development. This modularity allows for independent deployability, faster updates, and the flexibility for teams to choose the best tools and languages for each service, optimizing performance and productivity. Additionally, services can be scaled individually based on demand, promoting efficient resource usage. Lastly, this architecture also promotes fault isolation which means failure to one service does not impact the others, making the system resilient \cite{9779850}.

Microservice architecture promotes flexibility but it introduces several challenges. For example, one issue is effectively communicating the changes done in one service to other stakeholders. Breaking changes done in the API of one service can remain unidentified until runtime. Furthermore, development teams often rely on manual communication to notify other stakeholders about API changes which slows down the development process and increases the risk of human error \cite{10554880}. To address these issues, an API-First approach can be utilized which focuses on defining API before implementation allowing development teams to create API contracts that specify how a service will interact \cite{10.1007/978-3-642-39200-9_4}. Open API specification (OAS) can be used to define these API contracts making it easier for developers and machines to understand and interact with them. Lastly, one of the major benefits of OAS is to allow versioning of API contracts helping development teams to change and communicate the updates effectively.

While the Open API specification can provide significant advantages like versioning and effective communication but its adoption introduces certain challenges. Writing these specifications requires an understanding of the OAS format and tools related to it. Development teams might require training or they need to make themselves familiar with syntax and tools related to it so that the created specification document is both accurate and functional \cite{lazar2024specrawlergeneratingopenapispecifications}.   

Furthermore, once the API spec has been generated, development teams need to accurately translate it into working code which again increases the risk of human error. This problem can be resolved by using automatic code generators like Swagger Codegen \cite{ponelat2022designing}. However, this generated code might not follow the coding conventions of the organization and they need to make the necessary modifications. Also, this approach will primarily generate foundational elements such as request and response models, basic structural code, and documentation comments; the core functionality and business logic must be developed and implemented by the development team.

In order to address the above-mentioned challenges, introducing generative AI in the development process seems like a promising solution. AI-driven tools leveraging LLM can assist development teams in writing Open API specifications.  Such tools reduce the learning curve since the developer can generate the first draft of the specification simply by providing natural language prompts. Furthermore, LLMs can also be employed to generate more complex code compared to traditional code generators and according to the desired conventions, thus minimizing human error. However, one of the limitations of the current generation of LLMs is that they can produce only a few hundred lines of code at a time \cite{rasheed2024codeporilargescaleautonomoussoftware}. This limitation of LLMs is not a problem with microservice architecture because each service is independent, small, and focuses on a particular problem of the large system. Hence integrating LLMs can significantly boost developer's productivity.

This paper presents a system that helps in the creation of OpenAPI specifications and the generation of API code related to that specification. By integrating a chat interface, it allows developers to refine the generated code and specification through natural language prompts, easing the development process from design to fixing. The generated code will follow a predefined folder organization and also be deployable in the Docker environment. A key feature of this system is its ability to access the logs from the local development environment in order to guide and assist developers in debugging the service code. This access to logs improves the quality of fixes, as the system can offer more accurate, context-aware solutions. The system consolidates all these tasks under a single interface, significantly reducing the need for switching between multiple tools and sources. Developers no longer need to manually search through logs, error messages, and external documentation. Instead, they can rely on the system to gather necessary context, identify issues, and offer relevant fixes, making the development process more efficient. Moreover, this system utilizes a multi-agent workflow, where each agent is designed to perform a specific, well-defined task. For instance, one agent is responsible for generating the OpenAPI specification based on user input, another takes this specification to generate the server code, and yet another tests the code, identifies issues and suggests or applies fixes. This division of responsibilities ensures that the system is modular. Additionally, LLMs have limited memory, and by distributing tasks across multiple agents, the system minimizes the strain on any single agent. 

To validate the system's functionality and usability, we surveyed industry practitioners who had varying levels of experience in software development and microservice architecture. This allowed us to capture a broad spectrum of feedback on the system's usability and effectiveness across different expertise levels. Additionally, we made the data from our data analysis publicly available at \url{https://zenodo.org/records/14505669}, providing transparency and enabling further insights from the development community \cite{anonymouse_2024_14505669}.

\section{Background}
\label{Background}

\subsection{Generative AI}
\label{Generative AI}

Generative Artificial Intelligence (AI) is a branch of machine learning that can create realistic and complex data, like text or images, by learning patterns and structures from existing data \cite{10421601}. A wide range of fields such as technology, business, education, healthcare, and arts have been affected directly or indirectly by Generative AI \cite{10628898}, \cite{10589972}, \cite{rasheed2024can}. Even though it introduces some challenges like mode collapse, evaluation difficulties, ethical issues, and data quality it also offers diverse opportunities for amplifying creativity and productivity. There is a need for proper AI-human alliance so that these challenges can be eliminated and benefits can be maximized \cite{10556223}, \cite{9889652}, \cite{10421601}, \cite{rasheed2024timeless}.

Generative AI presents several technical challenges. One of the most common ones is hallucination, where it produces content that often looks correct but is factually incorrect or fabricated \cite{alkaissi2023artificial}. Depending on the industry, this can lead to serious consequences \cite{sallam2023chatgpt}. Incorporating Human-in-the-Loop (HITL) approaches can address concerns related to the accuracy of AI-generated content. By involving humans in tasks where precision is critical, the risk of hallucinations or incorrect outputs in the final product can be significantly reduced. Additionally, human feedback can provide rapid evaluations of generated content, helping the model to refine and improve the accuracy of its outputs over time \cite{christiano2017deep}, \cite{rasheed2025large}.

\subsection{Generative AI Models For Code Generation}
\label{GPT models in SE}

Generative AI models in simple terms are machine learning models that are trained using vast datasets giving them the ability to understand structures and patterns in data across different domains \cite{zhao2024surveylargelanguagemodels}, \cite{10661504}, . In recent years, these models have become a very popular choice for code generation tasks. These models combine natural language understanding with generative capabilities and have demonstrated exceptional performance in code synthesis. \cite {liu2023codegeneratedchatgptreally}, \cite{rasheed2023autonomous}, \cite{rasheed2024ai}. This has attracted the attention of many academic researchers and software developers \cite{jiang2024survey}. Another application of these code-generating models is code completion where code snippets are suggested based on partially written code. A more recent advancement in generative AI models for code-related tasks is the incorporation of function-calling capabilities in large language models (LLMs). Function calling allows models to execute structured API requests, interact with external systems, and automate complex workflows. Instead of merely generating code as output, function-calling LLMs can invoke predefined functions, retrieve real-time data, and even perform code execution \cite{kim2024llmcompilerparallelfunction}. This feature is particularly beneficial in software development, as it enables seamless integration with APIs, debugging tools, and deployment environments.
Despite many benefits and capabilities, Generative AI models in code generation tasks face some challenges. One of the most common and concerning ones is the quality of the code snippet generated, which often contains bugs or security vulnerabilities. As mentioned earlier, human involvement in the process of using LLMs to solve a particular problem becomes very crucial. Furthermore, fine-tuning the model on datasets containing vulnerability fixes can also address the security concerns in generated code \cite{wang2023enhancinglargelanguagemodels}.

\subsection{ API-First Approach in Microservices}
\label{API-First Approach}

\begin{figure}
    \centering
    \includegraphics[width=1\linewidth]{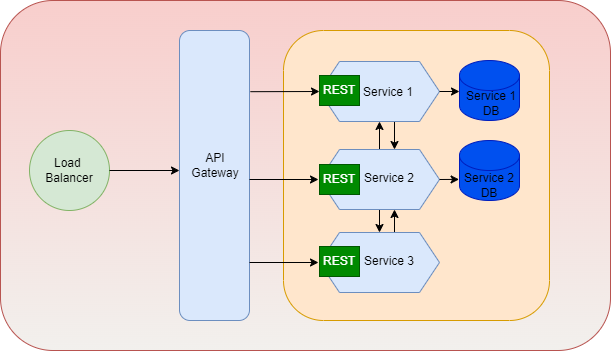}
    \caption{Microservice Architecture}
    \label{fig:enter-label}
\end{figure}

API-first or API-first approach is a design philosophy that prioritizes the design and defining of Application Programming Interfaces (APIs) before the implementation of its logic \cite{beaulieu2022api}. The API's inputs and output parameters are discussed first and defined concretely so that there is a clear understanding of the API's use cases and how communication with the API will take place. This leads to shared understanding and clear contracts among different services, which further promotes modularity since API clients and suppliers only need to follow the rules of these contracts. In other words, API definitions are treated as first-class citizens \cite{dudjak2020api}.

In microservice architecture, the system is first broken down into smaller independent services. Then API contracts of these independent services must be defined before the development team starts implementing them. Once interface specification has been established for each service it is shared among the team so that consumers and providers can work together. This leads to faster delivery and promotes re-usability. Depending upon the communication type i.e synchronous or asynchronous, suitable tools must be selected for writing the specification \cite{DudjakMario2020AAmf}.


To define these API contracts we must establish some standards so that we can ensure clarity and consistency which further improve collaboration. To achieve this standardization we can use OpenAPI specification (OAS) to define these API \cite{9650408}. OAS is a standard format for describing RESTful APIs, making them machine-readable and easy to share across teams and systems. It provides a complete definition of the API, including endpoints, operations, request/response formats, and security requirements. Moreover, the OAS definition can also be used to generate interactive documentation which can be used to interact with API and to grasp what to expect in the response \cite{9650408}.

\url{https://github.com/sirbh/sample_generated_cpi/blob/main/openapi_spec.yml}. Additionally, these specification documents can be versioned to track the changes, fostering clear communication about the updates done in the API, across various development teams or consumers.

As mentioned earlier API first approach treats APIs as first-class citizens and hence they can serve as a single source of truth. This enables development teams who are writing code for clients and servers to work in parallel, as a clear and well-defined contract has already been established. Furthermore, this concurrency also stretches to testing, allowing testers to plan test cases, and allowing client and server implementation to be validated more quickly. This synchronous development process allows development teams to detect faults early which further leads to rapid iteration promoting the agile development life cycle \cite{alma9911478695505973}.  

\section{Research Methodology}
In this section, we present the methodology for automating the entire development of
a service from generating Open API specification to testing and fixing the API code. Section \ref{RQs} provides details of the formulated Research Questions (RQs). The system design and multi-agent workflow are discussed in Section \ref{System Design} and we discuss the details of our evaluation framework in Section \ref{Empirical validation}.

\subsection{Research Questions (RQs)}
\label{RQs}
Based on our study goal, we formulated the following two Research Questions (RQs).

\begin{tcolorbox}[colback=green!2!white,colframe=black!75!black]
\textit{\textbf{RQ1.} How do users perceive the usability and effectiveness of a system that automates code generation and testing, compared to traditional manual coding methods?}
\end{tcolorbox}

The main objective of \textbf{RQ1.} is to assess user perceptions of the usability and effectiveness of an AI-driven system for code generation and refinement. This includes comparing the AI system with traditional manual coding methods regarding ease of use, efficiency, and user satisfaction. 

\begin{tcolorbox}[colback=green!2!white,colframe=black!75!black]
\textit{\textbf{RQ2.} To what extent does the system reduce the need for switching between multiple development tools in the development and testing of microservices?}
\end{tcolorbox}

The main objective of {\textbf{RQ2.} is to evaluate how effectively the system minimizes the need for switching between different development tools. This involves determining how much the AI system simplifies the process by integrating multiple functions, such as coding, testing, and deployment. 

\begin{tcolorbox}[colback=green!2!white,colframe=black!75!black]
\textit{\textbf{RQ3.}What impact does the system have on reducing manual coding efforts and increasing the speed of microservice development?}
\end{tcolorbox}

The main objective of {\textbf{RQ3.} is to analyze how the system reduces manual coding efforts and accelerates microservice development.

\subsection{System Design}
\label{System Design}

The system is scoped to the development of services that perform Create, Read, Update, and Delete (CRUD) operations and will communicate using REST architectural style, which is a perfect fit for these services because of its ability to handle CRUD operations using HTTP methods such as POST, GET, PUT and DELETE \cite{9320801}.
The system will use OpenAPI Specification standards to define the API specification as it provides a complete framework for defining REST APIs \cite{9650408}.
The LLM that is responsible for interacting with users is GPT-4 by OpenAI because of its great performance in generating accurate code snippets and interpreting complex natural language prompts. Moreover, its advanced function calling feature makes it a suitable model for this type of system \cite{10298721}. The system uses this function-calling feature of LLM to interact with the user's environment and to execute appropriate commands. 
The generated API code will be in JavaScript that uses the Express.js framework, a lightweight and flexible framework ideal for building RESTful web services. Unlike traditional code generators that primarily produce boilerplate code or a basic API skeleton, this system generates fully functional API code, including business logic. An example of a generated OpenAPI specification and its related code is provided here:  
\url{https://github.com/sirbh/sample_generated_cpi}

The system's workflow is quite structured and involves three stages as shown in figure \ref{spec_generation}, figure \ref{code_generation}, and figure \ref{server_interaction} which starts by taking input from the user and transitions into testing of generated server code. For this task, we have used a multi-agent architecture where each LLM agent has a distinct role. Details about the agents can be found in Table \ref{agents} which also contains details of functions that a particular agent can call in order to interact with the local environment of the user. 

The multi-agent architecture was adopted to enhance the system's modularity, flexibility, and scalability. By dividing the system's functionality into distinct agents—each responsible for a specific task such as code generation, testing, and specification generation — the system can operate more efficiently and be easily maintained.

Furthermore, the multi-agent architecture enables the replacement or updating of individual agents as needed. If there is a need to switch to a new language or framework, the code-generator agent can be replaced with one tailored for that environment. This eliminates the need to rework the entire system and makes it easier to update the technology stack without causing disruption.

In terms of memory management, the separation of responsibilities between the agents ensures that each agent maintains only the relevant context it needs. For example, the spec generator agent, which is responsible for generating the OpenAPI specification, does not need to be aware of code generation, fixes, or testing details. It simply handles the specification input and passes it to the appropriate agents. This clear separation allows for more efficient memory management, as each agent stores and processes only the information relevant to its task. By isolating memory usage between agents, the system reduces unnecessary data storage, avoids potential conflicts, and enhances overall performance.

Following are the steps involved in the process of generating, validating, and fixing the code for a service:


\begin{figure*}[t]
    \centering
    \includegraphics[width=0.7\textwidth]{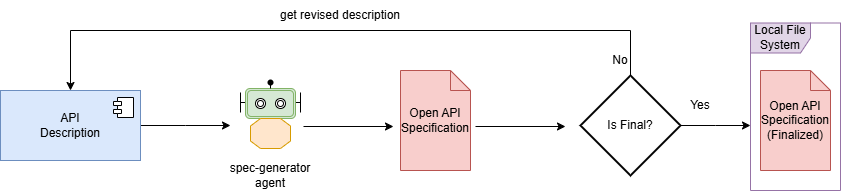} 
    \caption{Specification Generation}
    \label{spec_generation}
\end{figure*}

\begin{figure*}[t]
    \centering
    \includegraphics[width=0.8\textwidth]{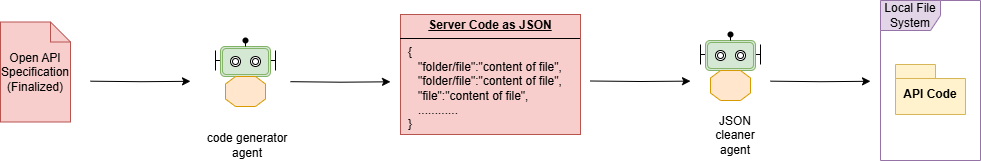} 
    \caption{Server Code Generation}
    \label{code_generation}
\end{figure*}

\begin{figure*}[t]
    \centering
    \includegraphics[width=0.7\textwidth]{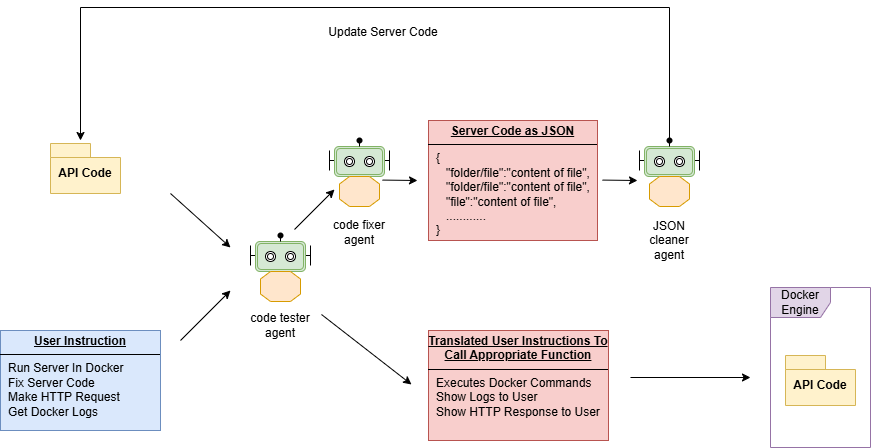} 
    \caption{Intracting With Server}
    \label{server_interaction}
\end{figure*}

\subsubsection{OpenAPI Specification Generation}

First as shown in figure \ref{spec_generation}, the user provides high-level requirements of the service which typically involve data models, endpoints, and type of database operation that are required to be implemented for example the following prompt can be given: \textit{generate OpenAPI specification for product service that can perform add product, edit product, delete product and fetches list of products operations. The product should have fields like name, description, price, and quantity}. Based on these inputs the system utilizes the GPT-4 model to generate the first version of OpenAPI specification. This task is mainly handled by the spec-generator agent. This step is very crucial in ensuring that the service interface is concretely defined and follow the OAS standards. An example of generated specification can be found here \url{https://github.com/sirbh/sample_generated_cpi/blob/main/openapi_spec.yml}

\subsubsection{Finalization of specification}

After the initial OpenAPI specification is generated, it is presented to the user for review. The user can make adjustments or provide additional details to the spec-generator agent to refine the specification. This iterative process continues until the specification accurately reflects the user's requirements. Once the user is satisfied by the output of the spec-generator agent, user can suggest to agent that spec is final as shown in figure \ref{spec_generation} and it can save it in the local environment of the user. Using GPT-4's function calling feature the agent will then call the necessary function which will take the specification in string format as input and save the specification in the user's local environment. This finalized version will serve as a blueprint or contract that will shape the rest of the development process of the service and hence will ensure consistency and purity as we move forward with the process.

\subsubsection{Server Code Generation}

The finalized OpenAPI specification of the service is provided to the Server Code Generator that contains two agents i.e. code-generator and JSON-cleaner as shown in figure \ref{code_generation}. The code-generator agent takes the specification and outputs a JSON string. It is prompted by the system with details such as the desired folder structure of the server code, the target programming language, and the framework in which the code should be generated. The structure of JSON is such that it represents the desired folder structure where the key represents the directory and the value represents the content of the file for example ``server/index.js'' will be the key and the content of file ``index.js'' will be the value. This JSON string may have some invalid tokens that may throw errors while parsing. To prevent the system from throwing an error the system, further passes this JSON string to a JSON-cleaner agent that cleans the JSON string so that it can be parsed without throwing any runtime error. To save the code in the local environment of the user, the JSON-cleaner agent calls a function that will parse the JSON, create directories and files based on the key, and fill those files with the content represented by the corresponding value. The final output after this step will be the saved server code in the user's local environment, organized with a logical folder structure. This includes all necessary files, such as the implemented server and business logic, database service for managing data, and configurations required to deploy the code in a Docker environment. An example of generated server code can be found here \url{https://github.com/sirbh/sample_generated_cpi/tree/main/express-server}

\begin{table*}[t]
    \centering
    \resizebox{\textwidth}{!}{ 
    \begin{tabular}{|>{\raggedright\arraybackslash}p{0.15\linewidth}|>{\raggedright\arraybackslash}p{0.2\linewidth}|>{\raggedright\arraybackslash}p{0.3\linewidth}|>{\raggedright\arraybackslash}p{0.25\linewidth}|} 
    \hline  
         Agent & Job & Functions Calls & Function Description \\ \hline  
         spec-generator & Generate OpenAPI specification from user API description & save\_openapi\_spec & Saves the given OpenAPI specification text to a YAML file and returns success/error. \\ \hline  
         code-generator & Generate code for server in JSON format and save it in user's working directory & save\_json & Validates and fixes a given JSON object before saving it as server code. \\ \hline  
         json-cleaner & Cleans JSON data of server files so that it can be parsed without error & & \\ \hline  
         code-fixer & Takes server code in JSON format with instructions for making fixes or updates and updates the code & save\_json & Validates and fixes a given JSON object before saving it as server code. \\ \hline  
         code-tester & Executes docker commands to start containers and fetch logs from containers. Also, send requests to service and show results to users. It is also responsible for making changes to the server code & 
         run\_docker\_compose \newline
         check\_docker\_compose\_status \newline
         get\_docker\_compose\_logs \newline
         run\_curl\_command \newline
         update\_json & 
         Start the services in the local docker engine \newline \newline
         Get the status of container-related to docker-compose file \newline \newline
         Get logs of containers \newline \newline
         Used to send HTTP requests to services \newline \newline
         Update/fix server code based on logs \\ \hline
    \end{tabular}
    }
    \caption{LLM Agent Details}
    \label{agents}
\end{table*}

\subsubsection{Automated Validation and Execution}
\label{Validation}
In order to interact with server code the system uses a code-tester agent. This agent heavily uses the function calling feature of GPT-4 LLM. The most important function it calls helps execute the ``docker-compose up.'' command that builds and loads the docker container in the user's local environment. For example user can give a prompt e.g. \textit{run docker containers} and the agent can use this prompt to match the most appropriate function out of the list of functions that are provided to this agent as shown in table \ref{agents}. It can then call the function that will execute \textit{docker compose up --build}.
Also, it can call functions that can help in fetching the logs of the container and the status of all the services running related to the server code if the prompts e.g.\textit{get logs related to service} or \textit{get service status} are provided. After calling the appropriate function it uses the returned data to show a summary of what's happening in the docker engine in a readable format. It eliminates the need to read the console logs which are often not user-friendly. Moreover, to validate the working of the server, the user can use natural language prompts e.g. \textit{get the list of products} or \textit{delete product with id}. The agent will then make requests to the server container by calling the function and getting the output to the user which eliminates the need to switch to any other tool like Postman to make requests and validate functionality. 

\subsubsection{Iterative Code Fixing}

After the user can interact with server code i.e. start and run containers, make requests to server containers, and get logs of running containers, there might be a scenario where the user encounters issues or something does not work as expected. In such cases, the code-tester agent, which has access to the server logs through its memory context if the user has asked for the logs (as discussed in previous steps), can assist. If the user asks the agent to detect the problem, the agent can analyze the logs and suggest potential fixes. Furthermore, it can interact with the code-fixer agent as shown in figure \ref{server_interaction} to modify the already saved server code by doing those fixes and restarting the services again to run the updated code. This eliminates the requirement of users to look through the code and make the updates, hence it can help in increasing productivity and saving time. To achieve this, the code tester agent prompts the code-fixer with the issue in code and server code in JSON format. The code-fixer agent then calls a function that takes two inputs i.e. the server code and the potential fix to a encountered problem. The output of the code-fixer is again a JSON string with keys as directories and values as the content of the files. This code is again parsed and saved, which updates the server code on the user's working environment. After the code is updated, the user can give a prompt to the code tester agent to rebuild and restart all the services and validate the server code, if it's working as expected. This functionality is repeated by keeping the user in the loop until the desired result is achieved.

\subsection{Evaluation Framework}
\label{Empirical validation}
To test the impact and capability of the system, a survey was conducted by getting direct feedback from industry practitioners. The tool's source code along with detailed usage instructions were provided to each participant. They were subsequently asked to complete a feedback form after using the tool to create a service.

The goal of this evaluation was to check and calculate how accurate the tools are in generating service code and how effective they are in debugging it. The system was first tested by each participant, after which an online survey was conducted to gather both quantitative and qualitative data.

\subsubsection{Questionnaire Design}
To effectively grade the system we carefully planned questions to capture both quantitative and qualitative insights from participants about their experience with the system.
The interview had two main sections:
\begin{enumerate}
    \item Participant's Background.
    \item Questions for tool evaluation.
\end{enumerate}

The survey begins with the questions about practitioner's overall experience with software development and with microservice architecture. The questions are in the Table \ref{background_questions}. These questions helped in making sense of the feedback based on their skill level.

\begin{table}
    \centering
    \begin{tabular}{|>{\centering\arraybackslash}p{0.05\linewidth}|>{\raggedright\arraybackslash}p{0.85\linewidth}|} \hline
        No. & Question \\ \hline
        1 & How many years of experience do you have in software development? Required to answer. Single choice. \\ \hline
        2 & How many years of experience do you have working with microservice architecture? Required to answer. Single choice. \\ \hline
        3 & Which programming languages are you most comfortable with? Required to answer. Open-ended. \\ \hline
    \end{tabular}
    \caption{Background Information Questions}
    \label{background_questions}
\end{table}

\begin{table*}[t]
    \centering
    \resizebox{\textwidth}{!}{ 
    \begin{tabular}{|>{\centering\arraybackslash}p{0.05\linewidth}|>{\raggedright\arraybackslash}p{0.7\linewidth}|>{\centering\arraybackslash}p{0.2\linewidth}|} 
    \hline
        No. & Question & Research Question Addressed \\ \hline
        1 & Did the tool successfully convert your natural language API description into an accurate OpenAPI specification? & RQ1 \\ \hline
        2 & How well did the generated server code align with the OpenAPI specification and your initial description? & RQ1, RQ3 \\ \hline
        3 & How much time did the tool save you compared to manually creating an API and server code? & RQ3 \\ \hline
        4 & How effective was the tool in helping you debug and fix issues in the generated code? & RQ1 \\ \hline
        5 & Did the tool’s testing features (e.g., sending API requests) effectively validate the functionality of the generated API? & RQ2 \\ \hline
        6 & If No, what was the issue? & RQ1, RQ3 \\ \hline
        7 & How effectively did the tool’s code update and fix features work in resolving identified problems? & RQ1 \\ \hline
        8 & How does using this tool compare to your usual method of API and server code development in terms of speed? & RQ3 \\ \hline
        9 & How does the code quality produced by the tool compare to what you would typically write manually? & RQ1, RQ3 \\ \hline
        10 & How would you rate the overall readability of the code generated by the tool? & RQ1 \\ \hline
        11 & Was the code organized in a logical manner (e.g., clear separation of concerns, modularity)? & RQ1, RQ3 \\ \hline
        12 & Do you feel the tool-generated code adheres to best practices and coding standards (e.g., naming conventions, formatting)? & RQ1, RQ3 \\ \hline
        13 & How likely are you to recommend the generated code to be used as part of a production system, purely based on its readability? & RQ1, RQ3 \\ \hline
        14 & How likely are you to replace or reduce your usage of other tools like Postman, Docker CLI, or your IDE if this tool offered comparable functionality in one interface? & RQ2 \\ \hline
        15 & What was the most significant advantage of using this tool? & RQ1, RQ2, RQ3 \\ \hline
        16 & What was the most significant disadvantage or limitation of the tool? & RQ1, RQ2, RQ3 \\ \hline
        17 & Would you recommend this tool to other developers? & RQ1, RQ3 \\ \hline
        18 & What improvements would you suggest for the tool? & RQ1, RQ2, RQ3 \\ \hline
        19 & How do you see this tool fitting into your regular development workflow? & RQ2, RQ3 \\ \hline
    \end{tabular}
    }
    \caption{Mapping of Survey Questions to Research Questions}
    \label{tab:survey_mapping}
\end{table*}

The core of the interview was about the effectiveness of the system where participants were asked to rate how well the system translated the natural language prompts into OpenAPI specification, alignment of generated server code with specification, and the effectiveness of the system in saving time, helping with fixing the code and validating the server code using natural language prompts. It was a mix of 13 single-choice and 6 open-ended questions. The single-choice questions include yes/no questions and statements with scale from 1 to 10, where 1 indicates strong disagreement and 10 indicates strong agreement. The open-ended questions focus on understanding developers' perceptions of the tool's advantages, limitations, and areas of improvement in their own words. 
This approach of using both open-ended and single-choice questions helped in ensuring that the feedback covered a wide range of aspects like code quality, folder structure, following best practices, and production readiness. 

\subsubsection{Practitioners Selections}
The participants were selected based on their experience in software development and familiarity with microservice architecture. A total of six developers participated in testing the system, representing a range of experience levels as shown in Table \ref{merged_experience_software} and Table \ref{merged_experience_microservices}. These participants were selected using convenience sampling.


\begin{table}[t] 
\centering
\resizebox{\columnwidth}{!}{ 
\begin{tabular}{|l|l|l|}
\hline
Experience Type           & Years of Experience                                                                                     & Number of Participants                                     \\ \hline
Software Development      & \begin{tabular}[c]{@{}l@{}}No Experience \\ 2--3 years\\ 4--5 years\\ 6--10 years\\ 10+ years\end{tabular} & \begin{tabular}[c]{@{}l@{}}1\\ 2 \\ 1\\ 1\\ 1\end{tabular} \\ \hline
\end{tabular}
}
\caption{Participants' Years of Experience in Software Development}
\label{merged_experience_software}
\end{table}


\begin{table}[t] 
\centering
\resizebox{\columnwidth}{!}{ 
\begin{tabular}{|l|l|l|}
\hline
Experience Type           & Years of Experience                                                                                     & Number of Participants                                     \\ \hline
Microservice Architecture & \begin{tabular}[c]{@{}l@{}}No Experience\\ 4-5 years\end{tabular}                                           & \begin{tabular}[c]{@{}l@{}}4 \\ 2\end{tabular}             \\ \hline
\end{tabular}
}
\caption{Participants' Years of Experience in Microservice Architecture}
\label{merged_experience_microservices}
\end{table}

Most of the participants had more than 2 to 3 years of software development experience, with one of them having more than 10 years of experience. Regarding experience with microservices, most participants had limited exposure, with 4 participants having no experience while 2 participants having 4-5 years of experience with microservices architecture.

By collecting real-world data from industry practitioners with different experience levels, the evaluation provided valuable insights into the tool's utility in real development environments. This approach helped identify both the strengths and areas for improvement of the tool. The smaller number of participants was intentional, as the tool is still in its early stages. This allowed for focused and in-depth feedback, ensuring that the data gathered was more actionable and directly applicable to the development process before expanding testing to a larger group.

\subsubsection{Data Collection}

The data collection process aimed to evaluate the system's usability and effectiveness based on participant feedback. The survey was conducted over a 2 month period, during which all six participants tested the system. Participants were emailed the usage instructions for the system \cite{saurabhcodegen}, along with a link to the survey form. They were instructed to use the system first and then complete the survey form to ensure their responses reflected their hands-on experience.

To facilitate structured testing, participants were provided with a detailed step-by-step guide outlining how to set up and interact with the system. The tool usage sessions were self-paced, allowing participants to work at their convenience. On average, it took participants 51 minutes and 16 seconds to fill out the survey.

\subsubsection{Data Analysis}

The data collected from the survey was analyzed using both quantitative and qualitative methods. Quantitative responses, which mostly included ratings on a scale 1 to 10, were assessed by calculating the mean to identify central trends and overall satisfaction levels across categories such as API accuracy, code quality, and time-saving ability. Modes were also calculated to identify the most frequent responses, that offered insights into the most common experiences participants had with the system. This helped to highlight recurring patterns in user feedback, providing a clearer picture of the system's overall performance. Qualitative data from open-ended questions was analyzed using thematic analysis to identify common patterns, suggestions, and feedback \cite{boyatzis1998transforming}. These approaches allowed a detailed understanding of both the numerical data and written text shared by participants.

\section{Preliminary Results}
\label{preliminary result}

\begin{figure}[t]
    \centering
    \includegraphics[width=0.4\textwidth]{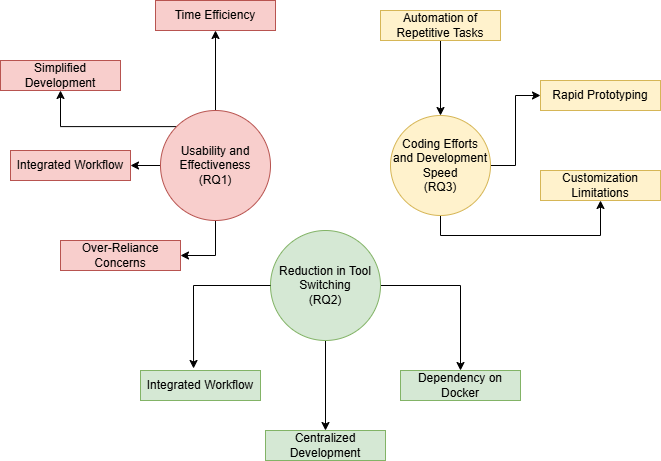} 
    \caption{Thematic Map}
    \label{themes}
\end{figure}

In this section, we present the results of data collected from the participants. The participants were assigned labels such as P1, P2, etc, and as shown in Figure \ref{data-analysis} their year of experience is labeled with their respective provided ratings for each attribute of the system. We will discuss about data collected for each attribute of the system in the sections below. Due to the small sample size, these results should be interpreted as indicative rather than definitive.


\begin{figure*}[t]
    \centering
    \includegraphics[width=1\textwidth]{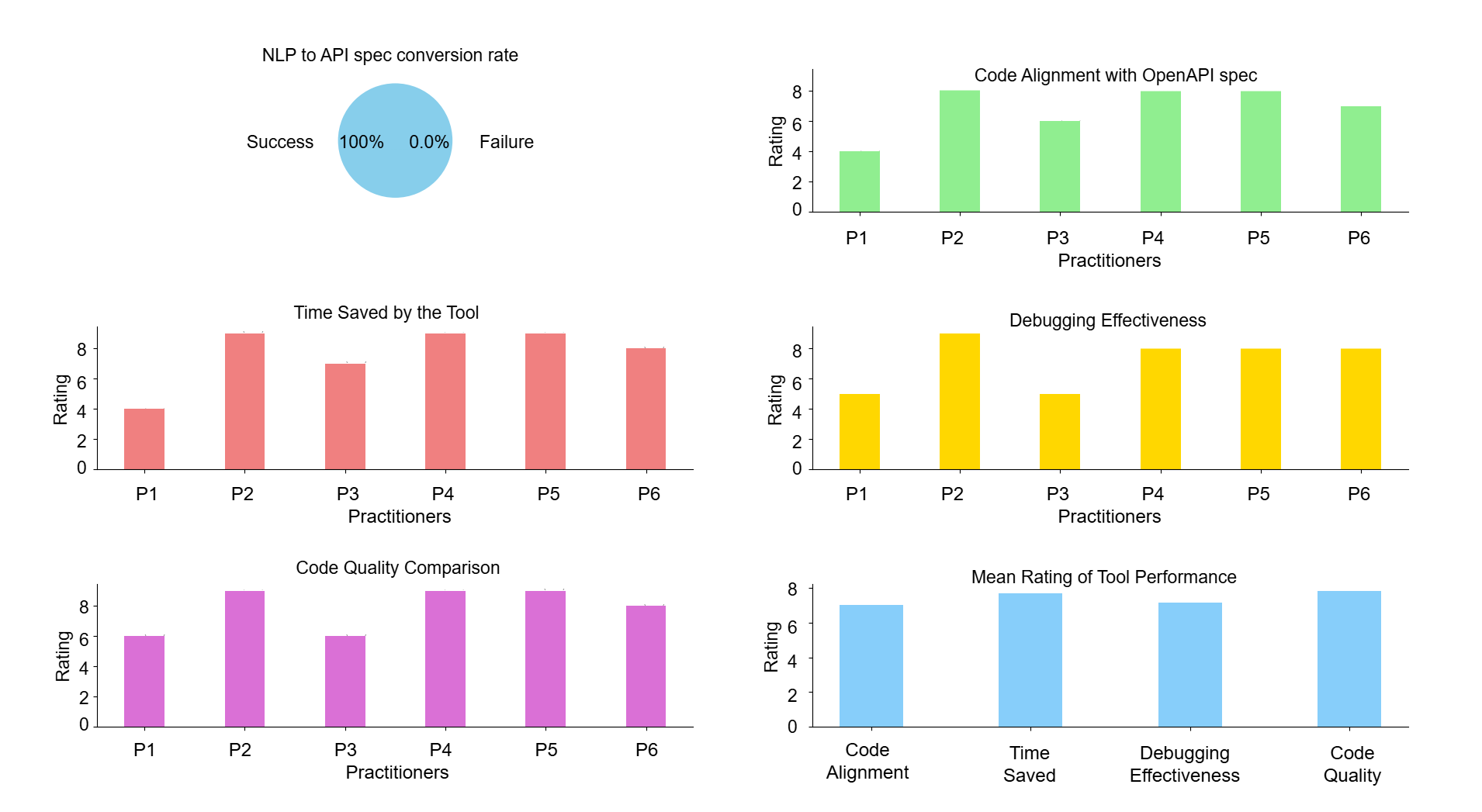} 
    \caption{Data Analysis Result}
    \label{data-analysis}
\end{figure*}

\subsection{RQ1: User Perceptions of Usability and Effectiveness}
Users generally perceived the AI-driven system as more usable and effective compared to traditional manual coding methods. The mean rating for the tool's speed compared to usual methods was 7.8 out of 10, indicating a significant perceived improvement in efficiency. Moreover, the system's ability to convert natural language API descriptions into OpenAPI specifications was positive, with all six participants reporting successful conversion. The mean rating for the alignment of generated server code with the OpenAPI specification was 7.0 out of 10 (mode = 8), indicating a high level of accuracy.
Thematic analysis of qualitative responses revealed three primary themes as shown in Figure \ref{themes}:\break\break
\textbf{\textit{Time Efficiency:}} Users find time-saving as a major advantage evident from survey question 3 in table \ref{feedback_questions} got a mean rating value of 7.67 out of 10. One respondent noted, ``It automates the whole process of building CRUD microservices, which saves a lot of time and effort.'' \break \break
\textbf{\textit{Simplify Development:}} Users appreciated the tool's ability to handle multiple aspects of development as the mean rating was 7 for survey question 14 in table \ref{feedback_questions}. One developer(P2) commented, ``It makes setting up CRUD microservices quick and easy by automating a lot of the boring, repetitive parts.'' \break \break
\textbf{\textit{Workflow Integration:}} Users saw the potential for the tool to streamline their development process. A developer(P4) commented, \textit{``I would definitely see it as a great 'starter-pack' for initial API Design and specification phase, rapid prototyping and MVP development with API-First approach in Agile Development.''}\break\break However, some users, particularly those with more experience, expressed concerns about over-reliance on the tool. One senior developer(P3) cautioned, \textit{``it is effective but can't rely totally.''}


\subsection{RQ2: Reduction in Tool Switching}
The system showed promise in reducing the need for switching between multiple development tools. The mean likelihood of replacing or reducing the usage of other tools was 7.0 out of 10 with the mode of 8, indicating a moderate to high potential for tool consolidation.
Qualitative responses supported this finding. \break\break
\textbf{\textit{Integrated Workflow:}} Users appreciated the tool's ability to combine multiple functions. A participant(P6) noted, ``The ability to interact with Docker, execute server code, and send API requests using natural language prompts eliminates the need to switch between multiple tools.'' \break\break
\textbf{\textit{Centralized Development:}} The tool's capacity to handle various aspects of development in one interface was frequently mentioned as an advantage. \break\break
\textbf{\textit{Dependency Concerns:}} One user(P2) expressed concern about the tool's reliance on Docker, which could be a limitation in certain development environments.\break\break

However, more experienced developers expressed some reservations. One user(P3) commented that while the tool is effective, they ``can't rely totally.'' on it, suggesting a need for integration with existing workflows rather than a complete replacement.

\subsection{RQ3: Impact on Manual Coding Efforts and Development Speed}
The system demonstrated a significant positive impact on reducing manual coding efforts and increasing development speed. The mean rating for time saved compared to manual creation was 7.7 out of 10, indicating substantial time savings.
Thematic analysis revealed two main themes:\break\break
\textbf{\textit{Automation of Repetitive Tasks:}} Users appreciated the automation of CRUD operations and API documentation. A user(P2) stated, ``It makes setting up CRUD microservices quick and easy by automating a lot of the boring, repetitive parts.'' \break\break
\textbf{\textit{Rapid Prototyping:}} The system was seen as particularly valuable for quick prototyping and MVP development. A developer(P4) noted its usefulness in ``rapid prototyping and MVP development with API-First approach..'' \break\break
However, some limitations were identified. A senior developer pointed out that for large-scale software development, the tool has limited customization and flexibility, ``particularly for complex features like custom authentication, authorization, data validation, error handling, and business logic.'' \break\break

\section{Discussion}

Several research efforts have been conducted to explore the use of LLMs for autonomous code generation \cite{rasheed2024codeporilargescaleautonomoussoftware}. However, these systems do not actually test the code in real-world environments, which can lead to potential issues such as configuration mismatches, missing dependencies, or platform-specific bugs that may not be caught during testing. To this end, our proposed system enhances this approach by directly executing the generated code within the user's local environment. By running the code locally, the proposed system ensures that the generated code functions as expected in the user's specific setup, addressing environment-specific issues right away. It also provides feedback based on logs generated and suggests fixes based on those logs, which helps the user resolve issues quickly. Unlike CodePori, which depends on abstract external verification done by LLM agents, this system offers a more reliable approach for validating code.

Other than this, traditional AI-assisted code generation tools, such as GitHub Copilot, offer significant improvements in developer productivity by providing intelligent code suggestions \cite{Zhang_2023}. However, these tools operate primarily in a static environment, relying on contextual information available within the codebase but lacking real-time feedback from the execution environment \cite{nguyen2022empirical}. This limitation often leads to inaccurate or incomplete code suggestions, as the model does not have access to runtime errors or deployment issues. In contrast, our system takes a more holistic approach by providing suggestions based on output logs in the runtime environment. 
Despite these advantages, there are potential challenges. The accuracy of error detection and fix suggestions depends on the LLM’s understanding of logs and its ability to interpret deployment-specific issues. Further improvements, such as fine-tuning models on domain-specific logs or integrating reinforcement learning-based corrections, could enhance the system’s reliability.

\section{Conclusion}

Participants involved in the interview, identified significant time savings, automation of repetitive tasks, and the ability to handle API design with code generation as the primary advantages. Participants with 2-3 years of experience highlighted the system's ability to automate the entire CRUD microservice generation process. Participants with more than 6 years of development found the system helpful in specification generation and documentation but suggested storing interaction history with the system. Another, suggestion from experienced participants was to improve the system's user interface.

Based on the feedback there are several directions in which the CRUD Microservice Code Generator can be expanded and improved. The most immediate focus is on incorporating open-source models to enhance flexibility and security. Integrating open-source alternatives to the current language model allows the tool to be adapted for a broader developer community and align more closely with open-source standards. Also, support for other languages and frameworks can be added like Python, Java, C\#, and other popular programming languages. This will allow developers from diverse backgrounds to leverage the tool within their preferred technology stacks.

Another key area for future development is to improve the system's ability to provide fixes because currently, it does require further refinement in certain cases where user intervention is necessary to resolve more complex issues in the server code. In some situations, the fixes suggested by the agents are not sufficient to fully address the problem, and additional manual adjustments are needed. To improve this, more advanced models can be explored, or the current LLM models can be further trained to handle these edge cases. This could enhance the system's ability to recognize and suggest better fixes and improve its understanding of different types of issues in server code. By refining the model’s capabilities to handle more complex scenarios autonomously, the system could reduce the need for user intervention and provide more effective solutions for code generation and debugging

Furthermore, the system's ability to fix and modify existing code while maintaining the desired folder structure can be expanded to larger projects by dividing the codebase into smaller chunks that can be managed by a single LLM agent. These agents can then work in teams to implement fixes or upgrades, improving efficiency and scalability. Additionally, more functions can be added to the LLM tool-calling feature to address more complex Docker commands or other system commands commonly used in development environments, such as Git commands or cloud deployment commands. This would allow users to interact with the system using natural language prompts, eliminating the need to remember complex commands. Instead of only providing input through text, a voice-based interface could also be developed to enable users to perform the entire development process, from generating code to deploying it, using spoken commands. This would further simplify the development workflow, making the tool more accessible to users.

To further validate and refine these improvements, a follow-up survey will be conducted with a larger number of developers after implementing the modifications. This will help in gathering more comprehensive data and insights, ensuring that the system evolves based on broader user feedback.

These advancements aim to increase the tool's adaptability and reduce the manual effort required in API and server code development, making it a more powerful and intelligent developer assistant.

\bibliographystyle{apalike}
{\small
\bibliography{example}}



\end{document}